\documentclass{article}
\usepackage{amsmath}
\usepackage{amssymb}
\usepackage[margin=1in]{geometry}
\usepackage[onehalfspacing]{setspace}
\usepackage[numbers,square]{natbib}
\usepackage{booktabs}
\usepackage{authblk}
\usepackage{hyperref}
\hypersetup{
	colorlinks=true,
	linkcolor=blue,
	filecolor=magenta,      
	urlcolor=cyan
}
\usepackage{ulem}
\usepackage{graphicx}

\title{Evolving higher-order synergies reveals a trade-off between stability and information integration capacity in complex systems.}

\author[1,2]{Thomas F. Varley}
\author[1,2]{Joshua Bongard}
\affil[1]{Departmenet of Computer Science, University of Vermont, Burlington, VT, USA}
\affil[2]{Vermont Complex Systems Center, University of Vermont, Burlington, VT, USA} 

\begin{document}
	
	\maketitle
	
	\begin{abstract}
		There has recently been an explosion of interest in how ``higher-order" structures emerge in complex systems comprised of many interacting elements (often called ``synergistic" information). This ``emergent" organization has been found in a variety of natural and artificial systems, although at present the field lacks a unified understanding of what the consequences of higher-order synergies and redundancies are for systems under study. Typical research treat the presence (or absence) of synergistic information as a dependent variable and report changes in the level of synergy in response to some change in the system. Here, we attempt to flip the script: rather than treating higher-order information as a dependent variable, we use evolutionary optimization to evolve boolean networks with significant higher-order redundancies, synergies, or statistical complexity. We then analyse these evolved populations of networks using established tools for characterizing discrete dynamics: the number of attractors, average transient length, and Derrida coefficient. We also assess the capacity of the systems to integrate information. We find that high-synergy systems are unstable and chaotic, but with a high capacity to integrate information. In contrast, evolved redundant systems are extremely stable, but have negligible capacity to integrate information. Finally, the complex systems that balance integration and segregation (known as Tononi-Sporns-Edelman complexity) show features of both chaosticity and stability, with a greater capacity to integrate information than the redundant systems while being more stable than the random and synergistic systems. We conclude that there may be a fundamental trade-off between the robustness of a systems dynamics and its capacity to integrate information (which inherently requires flexibility and sensitivity), and that certain kinds of complexity naturally balance this trade-off. 
	\end{abstract}
	
	\section{Introduction}
	
	Over the past two decades, information theory has emerged as something of a \textit{lingua franca} for the study of complex systems, as it provides a natural, mathematical framework for exploring the relationships between ``parts" and ``wholes" in multivariate systems \cite{varley_information_2023}. In particular, information theory can provide insights into ``emergent" or ``higher-order" interactions, where information is encoded in the interaction between large numbers of different elements \cite{varley_emergence_2022,mediano_greater_2022} and, crucially, not accessible from a reduced subset (challenging the historic focus on scientific reductionism). Newly developed formal tools like the partial information decomposition \cite{williams_nonnegative_2010,ince_partial_2017,varley_generalized_2023}, the integrated information decomposition \cite{mediano_integrated_2022}, and heuristics like the O-information \cite{rosas_quantifying_2019} have empowered scientists to start looking for higher-order information structures in a variety of complex systems, often with great success. Higher-order synergies appear to be ubiquitous in natural and artificial systems, having been found in climate data \cite{goodwell_temporal_2017}, sociological data \cite{varley_untangling_2022}, artificial neural networks \cite{tax_partial_2017, ehrlich_measure_2023, proca_synergistic_2022}, cortical neuronal networks \cite{newman_revealing_2022}, and global brain dynamics \cite{varley_multivariate_2023,varley_information_2023,luppi_synergistic_2022}. Furthermore, alterations in redundancy and synergy seem to reflect meaningful differences between systems. For example, synergy in on-going brain dynamics has been found to decrease when consciousness is lost \cite{luppi_synergistic_2023,luppi_reduced_2023}, and change with age \cite{gatica_high-order_2021,gatica_high-order_2022}. At the cellular level, the synergies instantiated by individual neurons changes depending on task performance \cite{varley_information-processing_2023} or drug administration \cite{varley_serotonergic_2023}, and in artificial neural networks, the distribution of redundancies and synergies changes over the course of the learning process \cite{tax_partial_2017,ehrlich_measure_2023}. 
	
	Despite these results, it remains unclear exactly what the significance of these alterations imply in the general case. Mediano et al., recently proposed that synergistic information integration can be understood as a general measure of ``computational complexity" in complex systems \cite{mediano_integrated_2022}, although ``complexity" is arguably as slippery a term as ``synergy" \cite{feldman_measures_1998}. Broadly, the relationship between higher-order information and other well-understood features of dynamical systems is relatively under-explored. 
	
	Generally, when scientists study higher-order information in a system, the presence (or absence) of higher-order information is the \textit{dependent} variable. How does synergy change when consciousness is lost? \cite{luppi_synergistic_2023,luppi_reduced_2023}? How does the synergy between social identities differ by demographic category \cite{varley_untangling_2022}? How do neural networks encode redundant and synergistic dependencies across training \cite{ehrlich_measure_2023}? In all of these studies, the change in redundancy or synergy is thought to be informative about some essential feature of the system's emergent properties, however synergy itself remains very abstract. 
	
	To try and resolve this ambiguity, in this study, we attempt to flip the script. Instead of treating higher-order information as the dependent variable that changes between conditions, here we ``force" particular kinds of higher-order information into simple boolean networks, and then characterize how the presence of higher-order redundancies, synergies, and complexities alters their dynamics. We choose boolean networks for several reasons. The first is that, as naturally discrete models, they are very amenable to simple information-theoretic analysis (indeed, elementary cellular automata have been a common test-bed for exploring emergent information dynamics \cite{lizier_local_2013,flecker_partial_2011,orio_dynamical_2023}). The second is that there is a long history of using boolean networks as toy models in complex systems. Seminal work by Kauffman showed how the structural properties of random boolean networks can inform on their dynamics \cite{kauffman_emergent_1984}, and we take direct inspiration from this work, although instead of altering structural properties of the network (degree, density, etc), we instead alter the computational properties of the individual nodes to inject redundancy or synergy into the logic. Finally, boolean networks are a very popular model in systems biology, where they are frequently used to model genetic and metabolic regulatory networks \cite{saadatpour_boolean_2013,albert_boolean_2014}.
	
	By exploring the link between the global information structure of a system and its dynamics, we hope to address two outstanding questions in the field of complex systems: the first is understanding what it means when we observe redundancy and synergy in natural and artificial systems (``what have we really learned upon discovering that a system is high in synergy"?), and the second to understand how systems that require particular attributes (canalization, computational capacity, etc) might self-organize their internal dynamics in a way that supports those particular properties. Resolving both of these questions will help scientists in many fields both better understand the structure and function of complex systems, as well as potentially helping with the design of novel systems with desirable computational or dynamic properties. 
	
	\section{Methods}
	
	\subsection{Boolean Networks}
	
	A boolean network is a directed graph: $\textbf{G}=\{\textbf{V},\textbf{E}\}$ composed of vertices \textbf{V} and directed edges \textbf{E} that connect two nodes $V_i\to V_j$. Each node $V_i\in\textbf{V}$ can be in one of two states: $\{0, 1\}$, and comes equipped with a function that maps the states of all the ``parent" nodes to an updated state of the target node: $f : \{0,1\}^k \mapsto \{0,1\}$, where $k$ is the in-degree of the target node. Being naturally discrete models, boolean networks have some useful properties for the purposes of this analysis: for a network with $N$ nodes, there will be only $2^N$ possible configurations the whole system can adopt (meaning a finite support set), and from any given initial condition, the system will always settle into an attractor after a finite period of time (meaning that the entire state-space can be brute-forced).  
	
	We used twelve-node boolean networks arranged into a ring lattice topology. Each node received four directed inputs from its four nearest neighbours, as well a self-loop ensuring that a given node's immediate past also factored into the computation of the next step (as is the case in the elementary cellular automata). A size of twelve was selected as it optimized the trade-off between the runtime required to compute the O-information (which grows with system size), and the richness of the state-space that could be plausibly explored. Unlike the standard elementary cellular automata (which have been frequently used in the past to explore higher-order statistics in discrete systems \cite{rosas_information-theoretic_2018,orio_dynamical_2023}), in our systems each node is allowed to implement its own unique function, vastly increasing the number of possible networks that can be evolved. 
	
	\subsection{Information-Theoretic Measures}
	
	We focused on three typical forms of higher-order information-sharing: redundancy (information duplicated over multiple elements simultaneously), synergy (information that is only present in the joint-state of all the variables and no simpler combination of sources), and ``complexity" (the balance between independence and integration \cite{tononi_measure_1994}). To estimate the redundancy and synergy, we used the O-information \cite{rosas_quantifying_2019}. First introduced by James and Crutchfield as the ``enigmatic information" \cite{james_anatomy_2011}, and then further refined by Rosas et al., the O-information provides a heuristic measure of whether a given multidimensional probability distribution ($\textbf{X}$) is dominated by redundancy (in which case $\Omega(\textbf{X})>0$ bit), or if it is dominated by synergy (in which case, $\Omega(\textbf{X})<0$ bit). Although the O-information can be written in several equivalent ways, here we prefer the form derived by Varley et al., \cite{varley_multivariate_2023} as it only requires defining one additional function:
	
	\begin{equation}
	\Omega(\textbf{X}) = (2-N)TC(\textbf{X}) + \sum_{i=1}^{N}TC(\textbf{X}^{-i})
	\end{equation} 
	
	Where $N$ is the number of elements, $\textbf{X}^{-i}$ is the set of all $X_j\in \textbf{X}$ \textit{excluding} $X_i$, and $TC(\textbf{X})$ is the \textit{total correlation} of \textbf{X}:
	
	\begin{equation}
	TC(\textbf{X}) = \bigg(\sum_{i=1}^NH(X_i)\bigg) - H(\textbf{X})
	\end{equation}
	
	The total correlation (and by extension, the O-information) is zero if $\textbf{X}$ is comprised of independent elements (i.e. $H(\textbf{X})$ is maximal). If we understand the total correlation in terms of deviation from independence, the O-information can be interpreted as a measure of whether \textbf{X}'s deviation from independence is in the ``whole" or the lower-order ``parts." If there is structure in the total correlation of the whole that is not accessible when considering the parts, then $(2-N)TC(\textbf{X}) > \sum_{i=1}^{N}TC(\textbf{X}^{-i})$. The reverse is true if most of the deviation from independence is in lower-order collections of elements. 
	
	To quantify the complexity, we used the Tononi-Sporns-Edelman (TSE) complexity \cite{tononi_measure_1994}. The TSE complexity was introduced by Tononi, Sporns, and Edelman in the context of theoretical neuroanatomy, and attempts to quantify the degree to which a system balances integration and segregation. A system is said to have a high complexity if, on average, each element is largely independent of every other element, but the whole system strongly deviates from independence. Like the O-information, the TSE complexity can be written out in terms of the total correlation:
	
	\begin{equation}
	TSE(\textbf{X}) = \sum_{i=1}^{N-1}\bigg[\bigg(\frac{i}{N}\bigg)TC(\textbf{X}) - \mathbb{E}[TC(\textbf{X}^{\gamma})]_{|\gamma|=i}\bigg]
	\end{equation}
	
	Where $\mathbb{E}[TC(\textbf{X}^{\gamma})]$ is the expected value of the total correlation of all subsets of $\textbf{X}$ with $\gamma$ elements. Since the number of subsets of size $\gamma$ grows with the binomial coefficient $\binom{12}{\gamma}$, it is not practical to sample all possible subsets for all scales of $i$. Consequently, following \cite{varley_multivariate_2023}, we took a subsampling approach: if the number of possible subsets of \textbf{X} was greater than 75, that was the number of randomly selected subsets sampled for each scale. 
	
	\subsection{Intervention Distributions}
	\label{sec:methods_maxent}
	
	\begin{figure}
		\centering
		\includegraphics[scale=0.5]{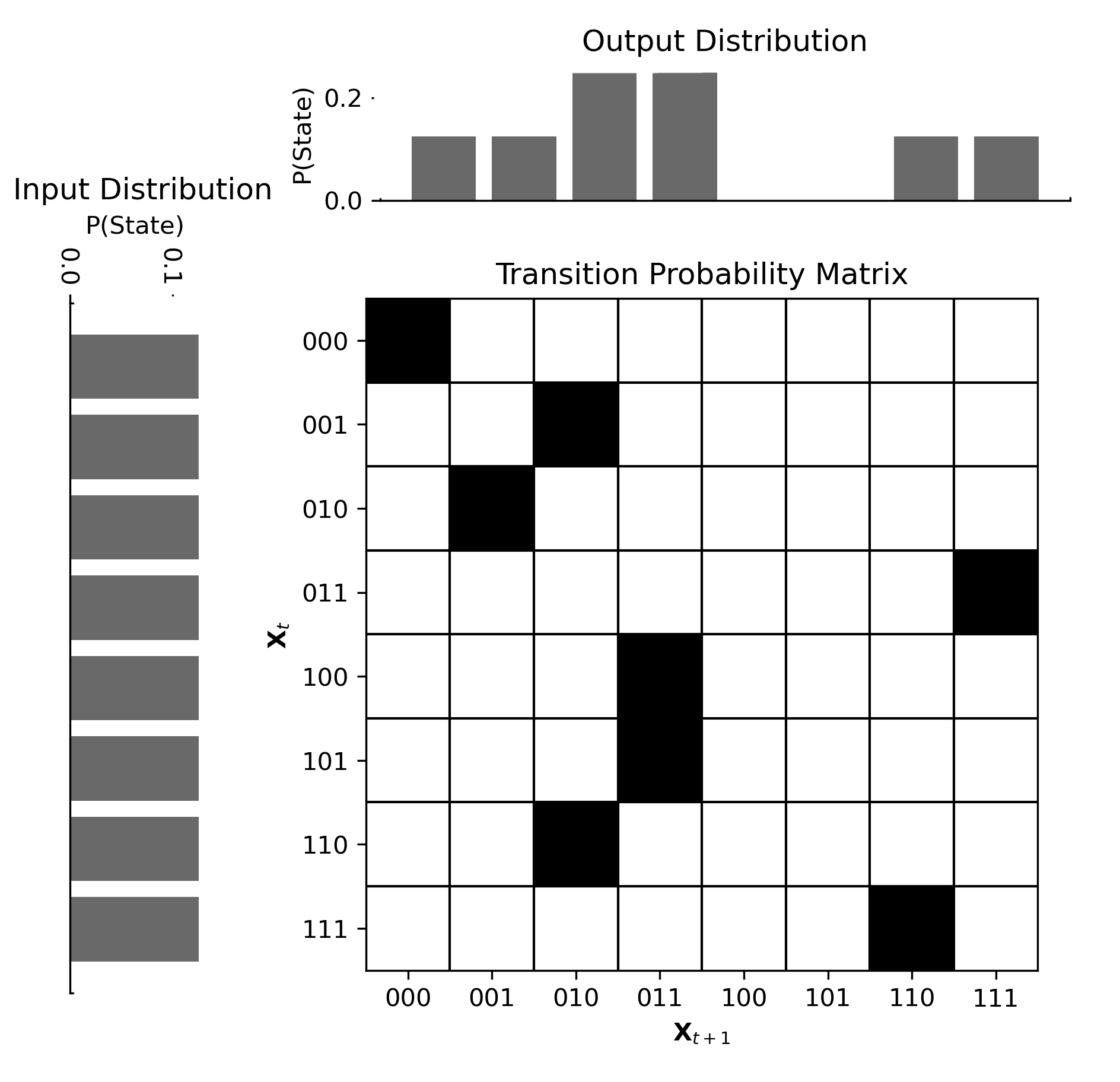}
		\caption{\textbf{Intervention distribution.} The state-transition structure of a boolean network defines a \textit{transition probability matrix} (TPM), which gives the probability of a given output (the columns), conditional on a given input (the rows). Here we see the TPM for a three-element system \textbf{X}, and how the intervention distribution is computed. At time $t$, all global states are equally likely, corresponding to a maximum-entropy distribution. After one timestep, to time $t+1$, the distribution of possible states is no longer uniform: the is the distribution of states after performing an intervention on \textbf{X} \cite{hoel_quantifying_2013} - it is this distribution that is fed into the calculations for O-information and Tononi-Sporns-Edelman complexity.  }
		\label{fig:tpm}
	\end{figure}
	
	Computing the information theoretic measures requires defining a probability distribution on $\textbf{X}$, which in this case is a distribution on the $2^{12}$ possible states a given boolean network could adopt. Here, we used a modified intervention distribution: initially, all possibles states $\textbf{X}=\textbf{x}$ were set to be equally likely (a maximum entropy distribution). Then, the system was allowed to update one timestep, producing a new distribution. Intuitively, this can be understood using a random walker model: if we imagine the state-transition structure of the whole system as a directed network with $2^{12}$ nodes, we can ``place" a single random walker on each node. Updating the system is like allowing the walkers to take a single step. The new distribution of walkers on nodes defines our intervention distribution: some nodes will now have more than one walker, some will have none, etc. 
	
	The use of the maximum entropy distribution can be understood in causal terms \cite{tononi_measuring_2003,hoel_quantifying_2013}: the updated distribution reflects the expected states of the system if it was intervened upon in a manner analogous to the \textit{do}-calculus \cite{hoel_quantifying_2013}. This causal flavour separates the intervention distribution from other plausible choices, such as the stationary distribution \cite{varley_emergence_2022}, which in determinsitic boolean networks, incorporates a very small subset of states. In the context of two-element systems, the use of this interventional distribution with the mutual information is known as the ``effective information" \cite{tononi_measuring_2003}, and so we can think about the application to the O-information or the TSE-complexity as the ``effective" O-information or ``effective" complexity as well. For a visual explanation of the intervention distribution, see Figure \ref{fig:tpm}

	\subsection{Evolutionary Optimization} 
	
	To evolve networks for redundancy, synergy, and complexity we implemented a naive evolutionary optimization. Initially, a population of 500 (if the measure was effective O-information) or 200 (if the measure was effective TSE complexity) random boolean networks were initialized. For each generation, the effective O-information or effective TSE complexity of each member of the population was computed, and the bottom half of the networks were removed. The remaining networks were then ``mated" to produce offspring, with pairs selected randomly, with the probability of being selected proportional to the rank of that network within the population. 
	
	The process of producing a ``child" network from two ``parent" networks is combining the functions associated with each node. Since all networks have the same topology, we matched the nodes in each parent and constructed a new function with 50\% of the input-output mappings coming from the first parent and the remaining input-output mappings coming from the second. We also built in an inherent mutation rate of 0.001, which is proportion of times a single input-output mapping is flipped. Each population was allowed to evolve for 750 generations, although the vast majority converged before 500 generations (for visualization, see Figure \ref{fig:evol_curves}).
	
	\subsection{Characterizing Boolean Network Dynamics}
	
	To explore how controlling higher-order information influences dynamics, we analysed the final, evolved systems with a suit of well-established tools used for exploring automata models. The first is the number of attractors that the system can settle in. Since deterministic boolean networks will always fall into either a point-attractor or a limit cycle, the number of attractors is an efficient heuristic for how canalized the network is (for more on canalization in boolean networks, see \cite{marques-pita_canalization_2013}). We also computed the transient time from every state to its inevitable attractor. Transient times are typically used to distinguish between boolean networks in an ordered, sub-critical phase (where transients are short) and a chaotic, super-critical phase (where transients are long) \cite{costa_effective_2023}. The final measure was the Derrida coefficient \cite{derrida_phase_1986}, which quantifies how robust the system is to perturbation. We followed the method detailed by Manicka et al., \cite{manicka_effective_2022}. Briefly, we selected 2000 possible states of the network, and randomly perturbed each state by flipping \textit{m} bits resulting in a state and a near-copy. Each state (original and copy) was then allowed to update one step and the Hamming distance between them was computed. We selected $m$ from the range $[1,2]$, and computed the Derrida coefficient as the slope of the linear regression of the average Hamming distance against $m$. A Derrida coefficient greater than one indicates a super-critical, chaotic system where small perturbations lead to large differences in future trajectories. A value less than one indicates a stable, sub-critical system where small perturbations have small effects. A Derrida coefficient of one indicates a critical system ``on the edge of chaos" \cite{langton_computation_1990}. These three measures provide a set of tools by which the relative stability or chaosticity of a given boolean network can be assessed and correlated with the presence, or absence, of higher-order information structures. 
	
	\subsection{Information Integration}
	
	To quantify the extent to which these systems are capable of integrating information (sometimes used as a proxy measure for non-trivial ``computation" \cite{newman_revealing_2022,mediano_integrated_2022}), we used a variation of the $\Phi$ value from integrated information theory \cite{balduzzi_integrated_2008} following the outline detailed by Mediano et al., in their recent paper \cite{mediano_integrated_2022}. Briefly, for a given boolean network \textbf{X}, we computed the difference between the effective information in the entire system and the sum of the effective informations after bi-partitioning \textbf{X} into non-overlapping subsets. This measure is typically referred to as $\Phi^{WMS}$:
	
	\begin{equation}
		\Phi^{WMS}(\textbf{X}) = I(\textbf{X}(t);\textbf{X}(t+1)) - \bigg(I(\textbf{X}^{\alpha}(t);\textbf{X}^{\alpha}(t+1)) +  I(\textbf{X}^{\beta}(t);\textbf{X}^{\beta}(t+1)) \bigg)
	\end{equation} 
	
	Where $\alpha$ and $\beta$ define the two, non-overlapping partitions of \textbf{X}. Ideally, the partition $(\alpha, \beta)$ is constructed to minimize the loss of effective information when partitioning the system (the so-called ``minimum information bipartition" \cite{tononi_measuring_2003}), however this is known to be an intractable problem for even modestly sized systems, as it required brute-forcing all possible bipartitions of \textbf{X}. Here, we use a heuristic estimator from spectral graph theory: the algebraic connectivity and the Fiedler vector \cite{fiedler_algebraic_1973,gross_handbook_2003}. 
	
	We began by constructing an effective connectivity graph of the system, defining a matrix \textbf{M} such that $\textbf{M}_{ij}=I(X_i(t);X_j(t+1)) + I(X_j(t);X_i(t+1))$. The result is a symmetric, low-dimensional representation of the dynamics of \textbf{X}, where each edge represents the total predictive information following from $X_i$ to $X_j$ and vice versa. For a dyadic, undirected graph, the, Fiedler vector is the eigenvector of the graph Laplacian associated with the smallest non-zero eigenvalue. From the Fiedler vector, it is possible to extract a bipartitioning of the graph that bisects the graph while approximately minimizing the total mass of the severed edges. This bisection defines the partition $(\alpha,\beta)$ used in computing $\Phi^{WMS}$. Spectral methods have been previously used to great effect in approximated the integrated information in large systems \cite{toker_information_2019}, and this method was chosen for its computational efficiency and conceptual simplicity (in contrast to other, more involved optimizations such as Queyranne’s algorithm that are sometimes used for this problem \cite{kitazono_efficient_2018}). To ensure that the effective connectivity graphs were connected, a small amount of noise was added to each edge, on the order of $10^{-6}$ bit. The computation of the Fiedler vector was done using the $\texttt{Networkx}$ package. 
	
	The $WMS$ in $\Phi^{WMS}$ refers to the idea that it is a ``whole-minus-sum": it is the difference in the predictive power of the whole system and the sum of the predictive power of the component parts. The higher the value, the more integrated information there is in the system. Unlike many information-theoretic quantities $\Phi^{WMS}$ is not strictly non-negative, and the interpretation of $\Phi^{WMS}<0$ was a standing question in the field for years. Recently, however, Mediano et al., showed that negative values occur when dynamic redundancy shared by $\textbf{X}^{\alpha}$ and $\textbf{X}^{\beta}$ overwhelms the integrated information \cite{mediano_towards_2021}. This prompted Mediano et al., to propose a corrected measure of integrated information: by adding back in the redundancy it is possible to ensure a non-negative value. Several different temporal redundancy functions have been proposed \cite{mediano_towards_2021,varley_decomposing_2023}, here we use the minimum mutual information redundancy measure following \cite{mediano_integrated_2022}:
	
	\begin{eqnarray}
	\Phi^{R}(\textbf{X}) := \Phi^{WMS}(\textbf{X}) + \min_{\gamma, \delta\in\{\alpha,\beta\}}I(\textbf{X}^{\gamma}(t) ; \textbf{X}^{\delta}(t+1))
	\end{eqnarray}
	
	We selected the top one hundred most redundant (highest O-information), most synergistic (lowest O-information), and most complex (highest TSE complexity) evolved boolean networks, and for each one we computed the minimum information bipartition and corrected integrated information $\Phi^{R}(\textbf{X})$. 
	
	\section{Results}
	
	\begin{figure}
		\centering
		\includegraphics[width=\textwidth]{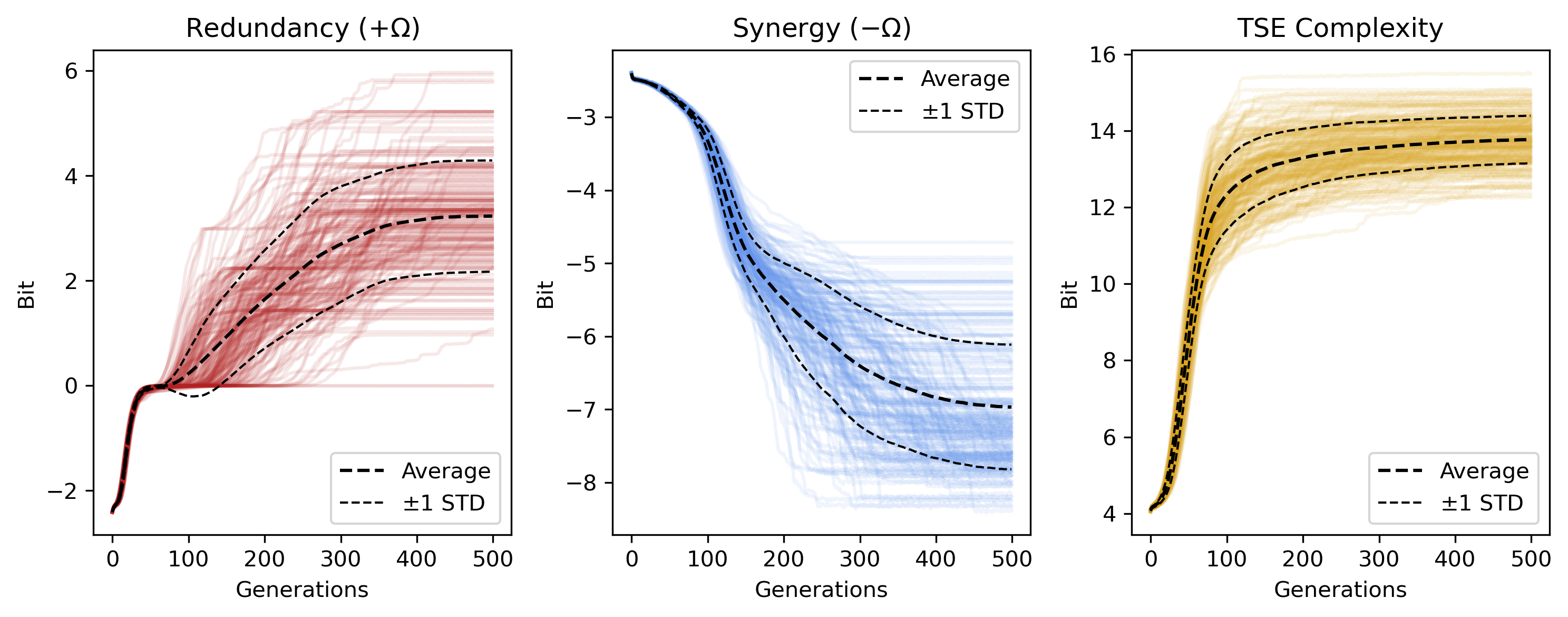}
		\includegraphics[width=\textwidth]{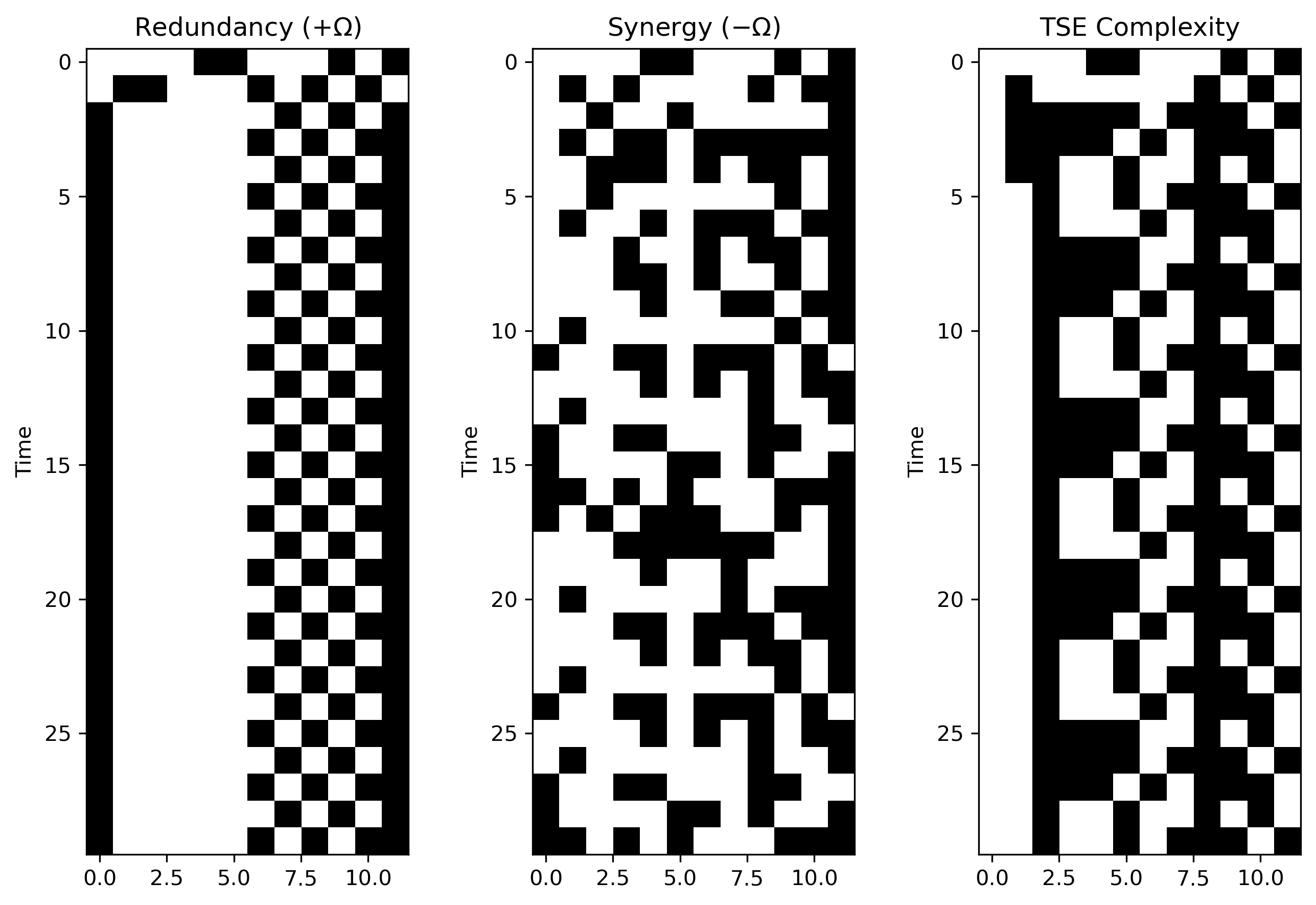}
		\caption{\textbf{Evolutionary optimization of redundancy, synergy, and complexity. Top row:} Presented above are the evolutionary trajectories for all populations evolving for redundancy (left), synergy (middle), and TSE complexity (right). From random initial conditions, it is clear that the evolutionary optimization is able to discover many configurations that significantly deviate from the random initial conditions. This figure establishes the colour schema that will be used for the paper. Red for redundancy, blue for synergy, and gold for TSE complexity. \textbf{Bottom row:} For each of the three classes of system (high redundancy, high synergy, and high complexity), we selected fittest boolean networks and ran them to get a visual sense of their different properties. Note how the highly-redundant system almost immediately falls into a stable attractor, while the high-synergy system has a long transient time and overall visually noisier patterns.  While these individual trajectories are one of many, they are representative of broader trends. For the redundant system, the average tansient time required to hit an attractor is merely 2.75 steps. In contrast, for the synergistic system, the average transient time is 26.91 steps, and for the TSE-maximizing system, the average transient time is 7.7 steps.}
		\label{fig:evol_curves}
	\end{figure}
	
	\begin{figure}
		\centering
		\includegraphics[width=0.75\textwidth]{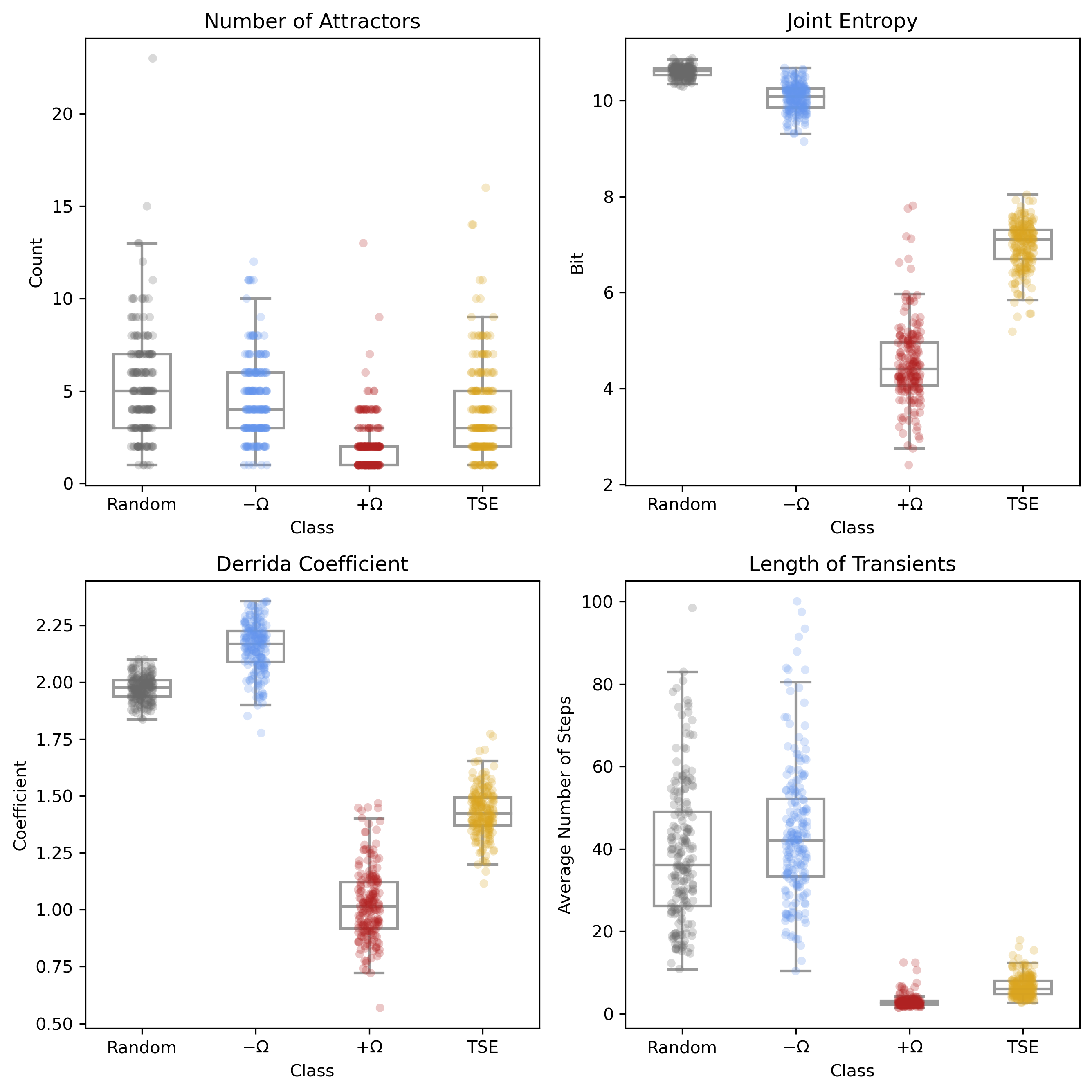}
		\includegraphics[scale=0.6]{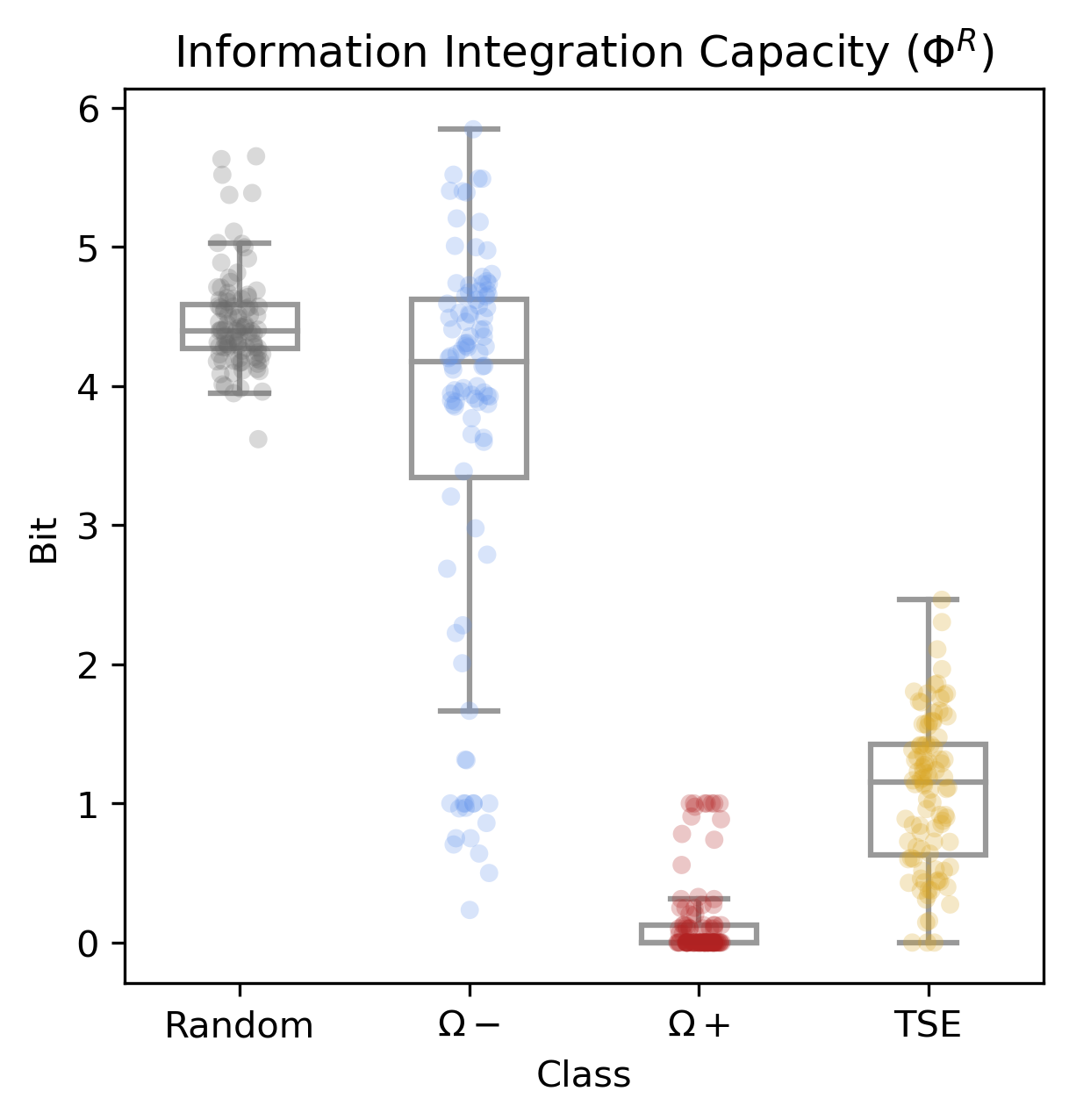}
		\caption{\textbf{Dynamical differences between random, synergistic, redundant, and complex systems. Top right:} The number of unique attractors for each network, for each class. \textbf{Top left:} The joint entropy of the global state intervention distribution after one timestep. \textbf{Middle left:} The Derrida coefficient for each network, for each class. \textbf{Middle right:} The average length of the transients. \textbf{Bottom:} The integrated information capacity $\Phi^{R}$ for each network for each class. }
		\label{fig:boxplots}
	\end{figure}
	
	The first result is verifying that the evolutionary optimization worked and successfully evolved a population with non-trivial higher-order information structures. Figure \ref{fig:evol_curves} shows this: for positive O-information, negative O-information, and high TSE complexity, we successfully evolved populations that significantly deviated from the random initial conditions. The average O-information for random systems was -2.396 $\pm$ 0.198 bit, while for the high-redundancy systems it grew to 3.246 $\pm$ 1.063 bit and for the high-synergy systems it fell to -7.02 $\pm$ 0.872 bit. Similarly, the random systems had a low TSE-complexity 4.107 $\pm$ 0.285 bit, but after evolving for high complexity, the systems achieved an average 13.919 $\pm$ 0.609 bit. The evolution for redundancy showed a curiously biphasic pattern: many of the populations transiently saturated around 0 bit, before accelerating above zero in a second phase of growth. 
	
	Collectively, these show that the evolutionary optimization works to construct systems that have desired higher-order information-structures, and also that there is an apparent link between "randomness" and synergy: randomly initialized systems already have O-informations significantly lower than zero, and larger random systems have lower O-informations. This link will be extensively explored in subsequent results. 
	
	\subsection{Network Dynamics: Stability and Chaos}
	
	When characterizing the different classes of systems, we found markedly different dynamics. The random systems had the highest average number of unique attractors ($5.285 \pm 2.715$), followed by very closely by the synergistic systems ($4.56 \pm 2.156$): this small difference was nevertheless significant ($U=16893.5$, $p=0.003$). The high-complexity systems came in third, with an average number of attractors of $3.91 \pm 2.607$, itself a significant decrease from the synergistic systems ($U=15494.5$, $p<4\times10^{-5}$). Finally, the redundant systems had the fewest average number distinct attractors ($2.13 \pm 1.457$), significantly lower than the high-complexity systems: $U=10314.0$, $p<3.77\times10^{-18}$. These results suggest that the redundancy has a ``canalizing" effect \cite{marques-pita_canalization_2013}: given that all networks have the same number of possible states ($2^{12}$), the low number of attractors in the redundant system indicates that more initial conditions can lead to the same final state than in the random or synergistic cases. 
	
	When comparing the joint entropy of the systems after maximum-entropy perturbation (see Section \ref{sec:methods_maxent}), we found that the random systems had the highest joint entropy ($10.606 \pm 0.126$ bit), which was significantly greater than the next-highest class: the high-synergy systems ($10.061 \pm 0.295$, $U=1469.0$, $p<4.08\times10^{-58}$). The high-complexity systems had the next-lowest joint entropy ($6.984 \pm 0.488$ bit), significantly lower than the high-synergy systems ($U=0.0$, $p<2.42\times10^{-67}$). Finally, the high-redunancy systems had the lowest joint entropy ($4.523 \pm 0.818$ bit), significantly lower than the high-complexity systems ($U=803.0$, $p<3.26\times10^{-62})$). These results are consistent with the notion that redundancy is canalizing while synergy (and randomness) maintain ``flatter" configuration landscapes, although unlike the number of attractors (which describes the limit behaviour of the systems), these show a collapse in joint entropy a mere single timestep after perturbation. This suggests dynamics that are not only canalizing in the limit, but rapidly canalizing in the short term as well. 
	
	We can see a similar result when we consider the time it takes the systems to reach their various attractor states (the transient times). The high-synergy systems had the longest transients ($43.965 \pm 16.405$ steps), which was just barely significantly greater than the random systems ($38.96 \pm 16.816$ steps, $U=16300.0$, $p=0.0007$). The high-complexity systems had radically shorter transient times ($6.671 \pm 2.634$ steps) compared to the random systems ($U=50.0$, $p<5.112\times10^{-67}$). Finally, the high-redundancy systems had very short transient times ($U=2146.5$, $p<4.27\times10^{-54}$). 
	
	The final measure of dynamic stability and chaosticity we measured was the Derrida coefficient \cite{derrida_phase_1986}, which quantifies the extent to which small perturbations propagate through time. A Derrida coefficient greater than one typically indicates chaos (i.e. small perturbations grow), while a Derrida coefficient less than one indicates stability (i.e. perturbations die out). We found that the synergistic systems had the highest average Derrida coefficients ($2.155 \pm 0.108$), indicating highly sensitive, chaotic dynamics. The random systems had the next-highest coefficients ($1.972 \pm 0.053$), a significant decrease from the synergistic systems ($U=3106.5$, $p<1.185\times10^{-48}$). The high-complexity systems were closer to the critical point ($1.428 \pm 0.103$), significantly lower than the random systems ($U=0.0$, $p<2.414\times10^{-67}$. Interestingly, the lowest coefficients were very near the critical boundary of one: the redundancy-dominated systems had average Derrida coefficients of $1.027 \pm 0.156$ (a significant decrease from the high-complexity systems, $U=1091.5$, $p<2.02\times10^{-60}$).
	
	Collectively, these results point to a consistent picture of how higher-order information can influence the dynamical properties of a boolean network: the presence of synergy seems to be destabilizing, producing signatures of chaos like sensitivity to perturbation, long transience, and lower canalization. In contrast, redundancy had a clear canalizing effect, although the Derrida coefficients also placed the redundant systems at the critical boundary between subcritical and supercritical dynamics. The significance of this is currently unclear. The high-complexity systems, which use TSE-complexity to balance integration and segregation seemed to split the difference: generally more flexible than the highly-redundant systems, but much more stable than the random or synergistic systems. For visualization of all results, see Figure \ref{fig:boxplots}.
	
	\subsection{Higher-Order Interactions Influence Information Integration Capacity}
	
	To assess how different higher-order information structures influenced the capacity of the system to integrated information, we selected the one hundred most redundant (highest effective O-information), most synergistic (lowest effective O-information), and most complex (highest effective TSE complexity) evolved boolean networks and computed the $\Phi^{R}$ measure of integrated information. We also generated one hundred random networks to compare the evolved systems to. 
	
	We found that the random networks had the highest value of $\Phi^{R}$ (4.46 bit $\pm$ 0.35). The second highest set was the synergistic set, with an average $\Phi^{R}$ value of 3.672 bit $\pm$ 1.45. The Mann-Whitney U test found significant differences between these groups ($U=6476.0$, $p=0.0003$). After the synergisitc systems, the family with the next-highest average $\Phi^{R}$ value were the high TSE complexity boolean networks (1.08 bit $\pm$ 0.537), which were significantly lower than the synergistic family ($U=1080.0$, $p<10^{-21}$). Finally, the redundancy-dominated networks had the lowest capacity to integrate information, with an average $\Phi^{R}$ of 0.162 bit $\pm$ 0.31. This was significantly lower than the TSE complexity group ($U=9308.5$, $p<10^{-25}$). For visualization see Figure \ref{fig:boxplots}. These results are consistent with the previous dynamic results: the highly synergistic systems resemble the random systems, while the highly redundant system is profoundly different and the high-complexity systems split the difference, appearing between the two extremes. It also suggests that redundant systems are not particularly effective at integrating information.
	
	\section{Discussion}
	
	After analysing a variety of signatures of dynamical complexity, we can begin to qualitatively describe the impacts of evolving for redundancy, synergy, and complexity. Broadly speaking, synergistic boolean networks resemble random boolean networks: both classes of network are highly sensitive to perturbation and have long transient times before reaching an attractor state (both indicators of chaotic dynamics). They also maintain high entropy dynamics and are capable of integrating information (as evidenced by the high values of $\Phi^{R}$). In contrast, highly redundant systems are typically stable, with fewer unique attractors, short transient times, and a general robustness to perturbation (suggesting sub-critical dynamics). They do not, however, integrate much information: their ``computational" or ``information processing" complexity \cite{mediano_integrated_2022} is very low. Finally, the systems that were evolved to have high TSE complexity \cite{tononi_measure_1994} seemed to sit between the redundant and the synergistic systems. They were more sensitive to perturbations and had slightly longer transients than the purely redundant systems, but also showed a greater capacity to integrate information. 
	
	These result suggest that there may be a fundamental trade-off between the capacity to integrate information, and the stability of the system. Systems that are too stable (such as those evolved for high O-information) cannot support information integration, while those systems that can integrate significant information are chaotic and unstable. This trade-off is reminiscent of the well-known compromise between redundancy and efficiency in the economics of supply chains: a supply chain with lots of redundancy can more effectively absorb shocks than a just-in-time chain that is more efficient but buckles under unexpected perturbations \cite{olson_trade-offs_2014}. Here, however, rather than a financial consideration, the consideration is whether the system is capable of ``integrating" information from multiple streams into a unified whole. 
	
	One possible implication of this trade-off is that, if one wanted to build a complex system that was both capable of integrating large amounts of information but was stable enough to be predictable (such as an animal nervous system), one possible avenue might be to expand the number of redundant elements enough to stabilize the information-integration capacity. This might partially explain the apparent upward pressure on the size of animal brains over the course of evolution, particularly the recent expansion of brain region associated with information integration \cite{luppi_synergistic_2022}: the capacity to engage in complex information-processing may require a ``substrate" of redundant components to stabilize the more complex integrative processes. This is, however, a significant leap from a small population of boolean networks and requires considerably more research before any strong claims can be made. 
	
	Maintaining a population of high-redundancy elements comes with its own energetic costs, and so a structure that innately balances the trade-off between stability and computational capacity may be still desirable. Our results suggest that a system with a high TSE complexity (which balances integration and segregation) \cite{tononi_measure_1994} may partially accomplish this. Across all measures, the systems evolved to maximize complexity (rather than high or low O-information) typically split the difference between the extremes of synergy and redundancy: being more stable than the high-synergy systems, but more flexible (and with greater computational capacity) than the purely redundant systems. These results are consonant with early findings by Sporns et al., who found that evolving for TSE complexity in simple systems produced topologies highly reminiscent of those seen in biological nervous systems \cite{sporns_theoretical_2002}. Future work further exploring the dynamical properties of the TSE-complexity, and the evolutionary contexts in which it emerges, may be highly informative about the general properties of evolved information-processing systems like the brain. 
	
	Finally, the apparently similar behaviour between synergy and randomness is an intriguing result. It is known that $\Omega(\textbf{X}) = 0\textnormal{ bit} \iff X_i \bot X_j$ for all $X_i$ and $X_j$ in \textbf{X} \cite{rosas_quantifying_2019}. The fact that the random systems had O-information values significantly less than zero strongly suggests a link between random systems and synergistic ones. This is not the first paper to suggest as such: Orio, Mediano, and Rosas recently showed that adding small amounts of stochastic noise to elementary cellular automata can transiently increase the synergy present in the system (as measured with the O-information as well) \cite{orio_dynamical_2023}, and Varley et al., argued that synergistic entropy (in the context of the partial entropy decomposition \cite{ince_partial_2017}) corresponds to irreducible randomness \cite{varley_partial_2023}. The nature of this think remains mysterious, as randomness is typically associated with independence (and by extension, minimal O-information). Where this "structure" comes from, then, is a question of significant interest. 
	
	\section{Conclusions}
	
	In this paper, we have shown how evolving small complex systems (in this case, twelve node boolean networks) for the presence of different higher-order information structures (redundancy, synergy, complexity) can influence the kinds of dynamics and computational capabilities the systems display. We found that evolving for redundancy produced systems that were robust to perturbation, highly canalized, and had low capacity to integrate information. Conversely, systems evolved for synergy resembled random systems in many respects: chaotic, sensitive to perturbation, but with a high capacity to integrate information. Finally, systems evolved to be TSE-complex (balancing integration and segregation) combined aspects of both extremes. We propose that there is a fundamental trade-off between stability and computational capacity, and that complex systems combining local segregation with global integration may naturally balance the two extremes.

\end{document}